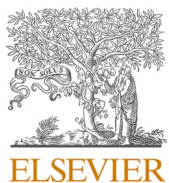

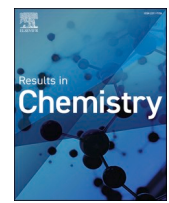

# Comparison of Gaussian process regression, partial least squares, random forest and support vector machines for a near infrared calibration of paracetamol samples

Aminata Sow [a, *], Issiaka Traore [a], Tidiane Diallo [b, c], Mohamed Traore [d], Abdramane Ba [a]

[a] Laboratoire d'Optique, de Spectroscopie et des Sciences Atmosphériques (LOSSA), Faculté des Sciences et Techniques (FST), Université des Sciences, des Techniques et des Technologies de Bamako, Bamako, Mali
[b] Département des Sciences du Médicament, Faculté de Pharmacie, Université des Sciences, des Techniques et des Technologies de Bamako, Bamako, Mali
[c] Laboratoire National de la Santé (LNS), Bamako, Mali
[d] Ecole Nationale d'Ingénieurs Abderhamane Baba Touré, Bamako, Mali

### ARTICLE INFO

### ABSTRACT

In this article, we analyze the near-infrared (NIR) spectra of fifty-eight (58) commercial tablets of 500 mg of paracetamol from different origins (that is, with different batch numbers) in the local markets in Bamako. The NIR spectra were recorded in the spectral range 930 nm-1700 nm. The samples are divided into forty-eight (48) samples forming the set of calibration (training set) and ten (10) samples used as the validation or test set. To perform multivariate calibration, we apply-three nonlinear regression techniques (Gaussian processes regression (GPR), Random Forest (RF), Support vector machine (KSVM)), along with the traditional linear partial least-squares regression (PLSR) to several data pretreatments of the 58 samples. The results show that the three nonlinear regression calibrations have better prediction performance than PLS as far as RMSE is concerned. To decide the best regression model, we avoid $R^2$ since this quantity is not a good parameter for this purpose. We will instead consider RMSE when comparing the different multivariate models. Additionally, to assess the impact of data preprocessing, we apply the above regression techniques to the original data, Multi-scattering correction (MSC), standard variate normalization (SNV) correction, smoothing correction, first derivative (FD), and second derivative correction (SD). The overall results reveal that Gaussian Processes Regression (GPR) applied to smooth correction gives the lowest RMSEP = 2.303053e-06 for validation (prediction) and RMSEC = 2.112316e-06 for calibration. In our investigation, one also notices that the developed GPR model is more accurate and exhibits enhanced behavior no matter which data preprocessing is used. All in all, GPR can be seen as an alternative powerful regression tool for NIR spectra of paracetamol samples. The statistical parameters of the proposed model are compared to the results of some other models reported in the literature.

## 1. Introduction

Amongst the drugs that have antipyretic and analgesic properties, paracetamol is almost the most prescribed analgesic drug nowadays. Counterfeiting this drug is then surely a moneymaking enterprise. Hence, the temptation to manufacture and sail fake paracetamol is very high. It is therefore an obligation to find mechanisms that allow easy and rapid identification of falsified or substandard paracetamol. One route is to be able to determine the drug's content. Traditional expensive approaches mostly destroy the drugs in the process of determination of their contents. Nevertheless, alternative nondestructive and low-cost methods which use near-infrared spectroscopy have now been very helpful in the front line of fighting against fraudulent medicines. In this new avenue, one is mainly interested in predicting the concentration (or content) of a chemical constituent in a sample from its near-infrared spectrum. These techniques of predicting concentrations (or contents) of drugs in terms of their NIR spectral data matrices in a suitable manner are known as multivariate calibration methods. Multivariate calibration techniques can be put into two categories: linear multivariate models and nonlinear multivariate models.

Some of the known linear multivariate models are for instance principal components regression (PCR), see for instance [1,2] and references therein, partial least-squares regression (PLSR) [2–4] and multivariate linear regression (MLR) [5,6]. An excellent overview of these linear calibration methods can be found in [7]. These types of regressions suppose that the content of a drug is linearly related to the spectral data matrices of that drug. The performance of this relationship is best measured on the independent test data using parameters such as the coefficient of regression $R^2$ as a metric and the root means square error of prediction (RMSEP). One interesting aspect of multivariate calibration is to achieve high accuracy (highest $R^2$ and lowest RMSE) with a minimum number of samples in the training set. This is important since it considerably reduces the number of laboratory tests and thereby reduces the cost of the analysis to be carried out in general. In most situations, the prediction of the contents of pharmaceutical medicines from their spectral data matrices using linear multivariate models faces problems due to the nonlinearity of multivariate spectral data matrices of medicines. To tackle this nonlinearity, the application of nonlinear multivariate calibrations to do prediction is very much in the air.

Noticeable nonlinear multivariate techniques which have been very useful in overcoming the problem of nonlinearity are support vector regression (SVR) [8], artificial neural networks (ANN) models [9], Gaussian processes regression (GPR) [10], Relevance Vector Machine (RVM) [11] and finally Random Forest (RF) [12,13]. These intelligent methods are used to perform classification [14,15]. For a recent review of these nonlinear techniques, see [16–18]. It has been shown that RF is a powerful tool for regression as far as NIR spectra are concerned [19]. Moreover, it also reported that Gaussian process Regression is more significant than Artificial Neural Networks [10]. What is more, it was demonstrated that GPR always out-performed PLSR, and SVM as far as RMSEV is concerned [20].

Because of all of these results, we seek the performance of GPR, SVM, RF and PLS in, this work. To save space, we have then judged it natural to avoid ANN and RVM in this paper. These two regression techniques will be considered elsewhere. Next, it is true that in doing regression analysis, caution must be taken when comparing various types of regression models. Since it has been proven that $R^2$ is a pitfall measure in judging for instance nonlinear models [21–25]. Therefore, in comparing the accuracy or goodness of the different calibration models we consider instead RMSE which has been used as a good measure in the field of NIR spectroscopy (see Figs. 1 and 2).

The aforementioned linear models, as well as nonlinear multivariate calibration models, have been applied to near-infrared spectral data of paracetamol in the literature [26–30]. It was observed that these techniques can accurately predict the content of paracetamol. What is more, predictions obtained by using nonlinear multivariate models are very much more accurate than those obtained by linear multivariate regression techniques. These results are first of all the source of motivation behind this investigation. Secondly, in our country, there is only one entity called LNS that does control quality (using traditional techniques) as far as drugs are concerned. It is then imperative to provide alternative methods in complement to techniques used at LNS. We apply NIR calibration apprs as such an alternative. Moreover, it is the first time to our knowledge that this analysis is applied to the NIR spectra of paracetamol samples sold at Bamako. It is hence mandatory to develop these nondestructive and low-cost techniques of identification for drugs sold in our region. For sure, masterregressionegressions techniques will lead to significantfightment in our fighting against adulterant drugs for the benefit of the patients. This article intends to bring a contribution to the application of regression techniques to paracetamol samples.

Another reason behind this study relies on the fact that in MALI,

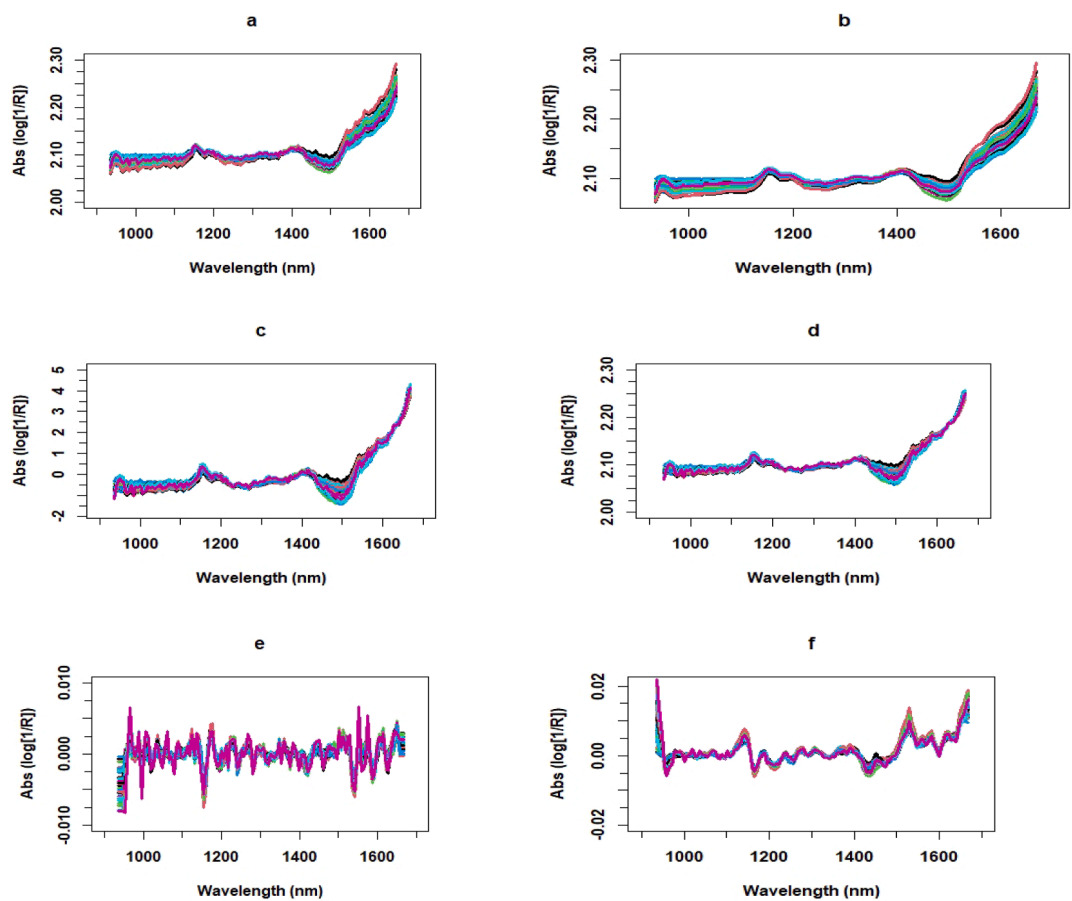

**Fig. 1.** a) the original spectra, b) the smooth spectra, c) the snv corrected spectra, d) the graph of msc correction, e) the plot of SD correction, and finally f) represents the plot of FD correction.

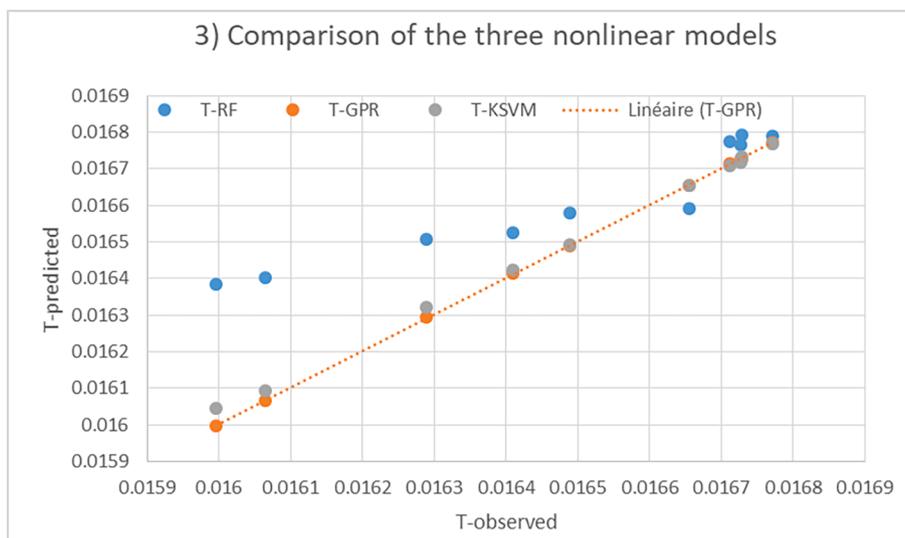

**Fig. 2.** 1) the prediction of GPR applied to smoothing, 2) the calibration of GPR applied to smoothing, and 3) the comparative plot of the predictions done by GPR, KSVM, and RF applied to smoothing correction.

there are several varieties of paracetamol depending on the industrical industries. What is more, paracetamol, unlike other analgesics such as aspirin, is well tolerated, i.e. has fewer side effects. In addition, it is available without a prescription, hence it is widely consumed. Consequently, there are many fake paracetamol drugs (problem with dosage, the total absence of the principle and out of date, etc.) on the Malian market,

All of these situations led us to study the quality of paracetamol through new regression methods such as GPR, SVM, RF and PLS. These techniques will allow us to analyze the quality of paracetamol so that to obtain paracetamol of good quality and lower risk for the population. To do so, we limited ourselves to fifty-eight samples randomly selected from 29 pharmacies in the district of Bamako. In each pharmacy, we collected two samples of paracetamol which added up to 58 samples of paracetamol. In the process of sampling, a given batch number is selected only once to avoid duplicates. So, we are working with fifty-eight different batch numbers of paracetamol. This should not be a problem in itself since for instance, the authors of [30] considered forty-five batch numbers of paracetamol.

The remaining parts of the application of this manuscript are sectioned as follows: in section two, we present the materials and the methodologies used for spectral data acquisitions. Section three is devoted to some basic theoretical regressionof the regressions techniques used in this work. The following section is devoted to results and discussion. The Conclusion is the heart of section five. We then finish this article by acknowledging the many supports we had while working on this project.

## 2. Materials and methods

### 2.1. Samples (Materials)

Fifty-eight (58) commercial Paracetamol 500 mg with different batches numbers were randomly purchased from 29 local pharmacies in Bamako. In the application of the regression models, these samples were divided into the training set and the test set. Besides, to obtain the contents of these drugs, we use a solution of NaOH with a concentration of 0.1n for the determination of the optical density. This thus allows us to determine the contents appropriately according to British pharmacopeia [31].

### 2.2. Methods

#### NIR data acquisition

In this study, we firstly labeled each of the fifty-eight tablets of paracetamol. To obtain a homogeneous compact powder of our samples, we triturated the tablets of paracetamol and passed them through a sieve with a diameter of 250 µm. We next weighted 0.206 g of each sample and used this quantity in our experimental setup. The NIR reflectance spectra of the samples are recorded with an optical flame-NIR-INTSMAS25 (800–1700 nm) spectrophotometer connected to a computer via a diffuse reflectance probe. For each sample, the spectrum is the average of 10 scans measured over the wavelength range 930 nm-1800 nm. This gives us a data matrix of 58 rows representing our samples and 128 columns labeling the wavelengths at which the reflectance of the samples has been measured. This matrix of reflectance is then transformed into a matrix of absorbance according to the well-known formula.

#### UV–VISIBLE data acquisition

The contents of our paracetamol are obtained by using a UV–visible spectrometer (Agilent Carry 630). The first step in this procedure allows us to obtain the optical density of our samples which is used in the computation of the content of the paracetamol. To achieve this measurement for a given sample, we again triturated 20 tablets of each sample. We next calculated the test portion (TP). We weighed each sample and diluted it with NaOH at a concentration of (0, 01 N). Finally, we used the aforementioned spectrometer and took the averages of ten 10 scans of every sample. This average yields the Optical density (OD) which is then used when computing the contents of the samples. The result is a matrix of 58 rows for the samples and 1 column for the contents (T). This is from now on denoted by $T(58 \times 1)$.

#### Data preprocessing

Before any analysis in NIR spectroscopy, it is advised and even sometimes mandatory to do some pretreatments of the NIR spectra. Different data pretreatments have different impacts and highlight different information about the data in question. In this paper, we apply to our matrices five types of data preprocessing. They are the standard normal variate (SNV) correction, the first derivative (FD), the second derivative (SD) which are obtained by using the Savitzky-Golay algorithm, the Multiplicative Scatter Correction (MSC), and lastly, the smoothing algorithm which is performed via the Savitzky-Golay algorithm (using the so-called *sgolayfilt* function). The description of these pretreatments is very standard and we do repeat them here. The pretreatments are done by using the R software.

To deal with regression of data in general, one has to divide the samples between the training set and the test set. Normally, the basic requirement is that 80 % of the samples should go to the training set and the remaining 20 % is set as the test set. Several mechanisms have been devised for solving this problem of partitioning the samples into these two sets. A pedagogical overview of some approaches which partially overcome the issue is described for instance in the package prospectr [32]. In this manusfirstwe firstly bypass this problem by making a combination of many techniques and selected 48 samples in our training set and 10 samples in the test set. The fifty-eight samples are then dived into a training set (48) and a test set (10).

## 3. Parameters of the different rmodelsn models

To properly apply a multivariate calibration model using the software R, one needs to specify some parameters before running the algorithm. In this section, the goal is not to go deeper into the meaning of these parameters but to instead simply pin down the values of some parameters used in this work. To perform PLS we use the PLS package [33], whereas KSVM and GPR are done by using the Package kernlab [34], and Caret [35]. We finally use the packages Random Forest [36] when performing RF. The package metric [37] is used in the computation of the bias of the different calibration models. In this article, we use the spectra of absorbance and calculate them from the measured Reflectance spectra $R(I \times J), I = 1, 2, 3, \cdots, 58 and J = 1, 2, 3, \cdots, 128$ of the different paracetamols I at the different wavelengths J, using the well-known relation.

$$A = X = Log_{10}\left(\frac{1}{R}\right) \quad (1)$$

### 2.1. Parameters of partial least square (PLS) algorithm.

The partial least square modelpresently onesently one of the most linear regression models used in NIR spectroscopy. This is a multivariate regression technique in which the algorithm uses the partial least square approach to find a relationship between the content $T[i]$ of a sample and the absorbance spectra $X[i, j]$ of this ample at different wavelengths. This is mathematically written as.

$$T[i] = T_0 + \sum_{j=1}^{128} B_j X[i, j] \quad (2)$$

Where $T_0$ is known as the intercept of the model, $B_j$ are the coefficients of the linear regression. PLS algorithm essentially yields computes the values of $T_0$ and $B_j$ in equation (2) and then simply uses them in predicting the content of new samples. To implement the algorithm using the software R,inituials to set some initials parameters. The first parameter is the number of components (ncomp) that must be used in modeling the PLS algorithm. In our investigation we set it to be ten (10) hence, (ncomp = 10). The next parameter that one must specify is the method of computation. For this, we choose the orthogonal scores pls (oscorespls) algorithm (aka the NIPALS algorithm). Finally, the last not the least parameter to be defined clearly is the validation type of the model. In our model building, we choose it as leave-one-out cross-validation (LOO).

For nonlinear regression types to be discussed below, equation (2) is fundamentally different. Nonetheless, the spirit is almost the same. The goal is to seek a relationship between the content of a sample in terms of its absorbance spectra measured at a different wavelength. To depart from equation (2) is then the main principle behind the different regression techniques. The theoretical starting details of them are beyond the scope of this investigation and we apologize for this missing.

We then move on by simply stating the parameters of the nonlinear calibration models.

### 2.2. Parameters of support vector Machine (KSVM) technique.

For this type of regression, one needs to specify the kernel parameter (kpar). This parameter kpar depends on the type of kernel used in the algorithm. We let the Algorithm automatically generates kpar. Hence through the computation, we choose kpar = automatic. This is very important, since by selecting a particular value of the parameters in kpar, one may dramatically alter the outcome of the regression. Besides, choosing kpar = automatic, the algorithm self-selects the best parameters when performing the regression. This algorithm necessitates assigning the type of problem one wants to do. Since it can be also used for classification. In this investigation, we opt for regression as the type of model building. In addition, the algorithm requires an identification of the cost of constraints violation (C) which is the 'C'-constant of the regularization term in the Lagrange formulation of the problem, here we take C = 4 different from the default value (C = 1). One further must give a value to the tolerance of the termination criterion (tol) and the kernel, in our work tol = Last but not least, we have to specify the kernel used in the regression. We selected it to be the **Linear (vanilla) kernel function.** These are the most useful parameters we selected in building our KSVM regression models.

### 2.3. Parameters of Gaussian process regression (GPR) algorithm.

We do not need to set many parameters for this algorithm. Surely, one can still personalize as many as possible parameters when running the algorithm. However, we do not do this as far as we are concerned, we simply specify the initial noise variance to be var = 0.2, and we again set kpar = automatic. The algorithm is run with the polynomial kernel function (polydot) as the kernel of the GPR regression. Moreover, in the algorithm of GPR, we adopted a 5-fold cross-validation approach. We observed that changing the number of cross-validation, such as to 10-fold, does not considerably change the outcomes of the computation.

### 2.4. Parameters of random forest (RF) algorithm.

To properly select the parameters in this regression, we first tune the number of predictors sampled for splitting at each node. This parameter is known as **mtry**. The best value of mtry is obtained by plotting the **Out-of-bag (OOB)** error, also known as the **out-of-bag estimate** against mtry**. The plots for the different pretreatment are given below. Mostly, these set mtry = 36. The next parameter to be specified is the number of trees denoted by **ntree**. This is the number of trees to grow in the forest. The basic requirement (to ensure that every input row gets predicted at least a few times by the algorithm) is that it should not be set to too small a number. We can fantastically get an idea of it by plotting the random forest which is the plot of the error against ntree. This gives a range of choices of ntree. The final important parameter is the **node size**. This is the minimum size of terminal nodes. The fact is that choosing a larger value for this number causes smaller trees to be grown. This then reduces the time of running the algorithm. Although this may sound perfect, it is not a synonym for the goodness of the prediction made by the model. We should stress that the default values are different for classification (node size = 1) and regression (node size = 5). In our work, nevertheless, we set node size = 3.

## 4. Results and discussion

In this section, we summarize and discuss the different results found in our investigation. It is important to stress that, some results will be omitted to save space. These are not going to affect the conclusion of our investigation.

In judging the performance of a regression technique, it is widely advised to consider some statistical parameters as guiding principles. We compare the different models constructed from the various data pre-processing by computing the multiple correlation coefficient $R^2$, the Root Mean Square Error $RMSE$ and the bias which are given by the following equations (3)–(5) below. The best model is characterized by the smallest values of $RMSE$, (very) good value of $R^2$ and also unbiased as possible.

$$R^2 = 1 - \left( \frac{\sum_{i=1}^{N} \left(T_p[i] - T[i]\right)^2}{\sqrt{\sum_{i=1}^{N} \left(T[i] - \overline{T}\right)^2}} \right) \tag{3}$$

$$RMSE = \sqrt{\frac{\sum_{i=1}^{N} \left(T_p[i] - T[i]\right)^2}{N}} \tag{4}$$

$$bias(\%) = \left[ \sum_{i=1}^{N} \frac{T[i] - T_p[i]}{N} \right] \tag{5}$$

Where $T[i]$, $\overline{T}$ and $T_p[i]$ represents the observed value of the content, its mean, and the predicted value of the content, respectively. N = 10 for validation and N = 48 for calibration. The results of the computation are given in Table 1 below.

The first results we would like to discuss are the statistical parameters from the regression models. These are in Table 1 and Table 2 below.

One witnessed the conclusion of Ref. [20] advocating that the Gaussian-outperformed-performed the other regression models. At the next level of comparison, it is visible that the Kernel Support Vector Machine or support vector machine (KSVM) is a very powerful tool that can be used after the Gaussian process. It is followed by the Random Forest approach as a conclusion from Table 1. Yet first derivative spectra is important as stated by the authors of [26], but by scrutinizing the values in Table 1, we conclude that the performance of GPR applied to smooth correction is more encouraging and give better result compared to the other models. This is also supported by the conclusion from consideration of KSVM. This conclusion is not fundamentally different from that of Ref. [26], because the authors of that paper have neither considered Gaussian process regression nor random forest. Another remarkable conclusion from Table 1 is that Gaussian process regression, as well as partial least square regression, are almost unbiased as far as the calibration set is concerned. The last conclusion but not the least is the fact that Random Forest and partial least square give more enhance results with the preprocessing SNV. This somehow indicates that the best model for RF and PLS is obtained when the SNV pretreatment is used.

From Table 2 and Table 1, one observes that the number of support vectors together with the objective function value determines the usefulness of the data preprocessing as far as KSVM is concerned. Additionally, we see that the value of cross-validation error fully gives the necessary information about the GPR regression. In the sense that the less this value for pretreatment, the better is GPR result at this preprocessing. Surely, one may avoid cross-validation, but still, we observed that it is important to do it when applying GPR regression to NIR spectra.

We now turn to the results of the graphs from some pretreatments and some regression models. To set the tone, let us begin with the ones from data preprocessing and see how they affect the original spectra.

One can see the impact of the different corrections to the original data. It is visible that SNV and MSC corrections significantly affect the original spectra in the spectral range 1500 nm- 1700 nm. This is surely the signal of the presence of noise in this range. We see that these two corrections are very much alike. Supporting the conclusion of [38]. The preprocessing techniques handle many unwanted features that one may face during data analysis. In oticle, we select the aforementioned data preprocessing due to the type of problem we are looking at, i.e., multivariate regression.

The random forest technique forces one to choose the number of trees to properly run the algorithm. One important device is to plot the **out-of-bag estimate** against mtry. The following graphs represent two of them representing the random forest of smoothing correction and the original data.

The right-hand side is for the smoothing correction whereas the left-hand side depicts the original data.

Next, let's finish up this section with the graphs of some regression

**Table 1**
Summary of the different regression models used in our investigation.

| Treatment | Regression | Calibration | | | Validation | | |
|-----------|-----------|-------------|---------|------|------------|------|------|
| | | $R^2$ | RMSEC | Bias | $R^2$ | RMSE | Bias |
| **Original** | **GPR** | **0.9999853** | **2.315791e-06** | **−4.336809e-19** | **0.9988283** | **1.184846e-05** | **−7.071851e-06** |
| | KSVM | 0.9967178 | 3.645706e-05 | −4.969783e-07 | 0.997501 | 2.905556e-05 | −2.566343e-05 |
| | RF | 0.9793752 | 0.0001151649 | 2.221332e-06 | 0.9029424 | 0.0001619364 | −0.0001064913 |
| | PLS | 0.9197801 | 1.169680e-04 | −4.091056e-17 | 0.7752049 | 0.0002539404 | −0.00000038 |
| **SNV** | **GPR** | **0.9999848** | **2.302214e-06** | **1.445602e-19** | **0.9965151** | **7.708547e-05** | **1.982052e-05** |
| | KSVM | 0.9956287 | 3.762566e-05 | 5.895032e-06 | 0.9845673 | 4.241855e-05 | −1.532585e-05 |
| | RF | 0.9640272 | 0.0001244175 | −2.015218e-05 | 0.9652768 | 0.0001660171 | −6.448813e-05 |
| | PLS | 0.9313926 | 0.0001081712 | −2.168404e-19 | 0.8779796 | 0.0003070678 | −0.0002441669 |
| **MSC** | **GPR** | **0.9999845** | **2.312966e-06** | **5.059611e-19** | **0.9963789** | **2.009719e-05** | **−1.101322e-05** |
| | KSVM | 0.9958699 | 3.721949e-05 | 7.706465e-06 | 0.9885027 | 3.198883e-05 | −8.978984e-06 |
| | RF | 0.9839591 | 0.0001151889 | −5.115568e-06 | 0.9217296 | 0.0001667694 | −0.0001193549 |
| | PLS | 0.9387181 | 1.022332e-04 | 1.662443e-17 | 0.7829787 | 0.0003242548 | −0.0002496721 |
| **SMOOTH** | **GPR** | **0.9999915** | **2.112316e-06** | **−7.011174e-18** | **0.9999659** | **2.303053e-06** | **−1.496828e-06** |
| | KSVM | 0.9967769 | 3.410321e-05 | −6.514476e-07 | 0.9993306 | 2.192707e-05 | −1.184869e-05 |
| | RF | 0.9850786 | 0.0001140908 | −2.780361e-07 | 0.9022344 | 0.0001867039 | −0.0001277817 |
| | PLS | 0.6153732 | 0.0002561214 | 5.449923e-17 | 0.6562349 | 0.0003054578 | −0.0001760158 |
| **SD** | **GPR** | **0.9999857** | **2.066435e-06** | **−2.891206e-19** | **0.917449** | **0.0001071367** | **−6.64164e-05** |
| | KSVM | 0.9763265 | 7.968738e-05 | −2.670681e-06 | 0.8768164 | 0.0001575389 | −0.0001252967 |
| | RF | 0.9856274 | 0.00100464 | −1.414325e-05 | 0.9209222 | 0.000203555 | −0.0001584418 |
| | PLS | 0.9623445 | 8.013839e-05 | −7.950823e-19 | 0.3855253 | 0.0003550304 | −0.0002650413 |
| **FD** | **GPR** | **0.9999894** | **1.957563e-06** | **7.22802e-20** | **0.9581358** | **7.82742e-05** | **−5.49845e-05** |
| | KSVM | 0.9938056 | 4.239372e-05 | 2.636091e-06 | 0.9093767 | 0.0001051348 | −5.580012e-05 |
| | RF | 0.9636043 | 0.0001256483 | −8.042707e-06 | 0.9368729 | 0.00019699 | −0.0001394919 |
| | PLS | 0.9512478 | 9.118492e-05 | −8.673617e-19 | 0.7179846 | 0.0003150645 | −0.0002287434 |

The outcomes of the GPR and KSVM regressions give some more parameters such as training error, cross-validation error (GPR only), number of support vectors (NS-Vectors), and finally the objective function Value (OF-Value) (KSVM). These parameters for the different pretreatments are given in Table 2 below.

**Table 2**
Some Specific parameters of the regression models (GPR and KSVM).

| Regression | Parameter | SNV | SMOOTH | MSC | FD | SD | Original |
|---|---|---|---|---|---|---|---|
| GPR | Cross-V Error | 5e-09 | 0 | 5e-09 | 3.5e-08 | 4.1e-08 | 1e-09 |
| | Training Error | 3.043e-05 | 2.5617e-05 | 3.0714e-05 | 2.2001e-05 | 2.4516e-05 | 3.079e-05 |
| KSVM | Training Error | 0.008128 | 0.006677 | 0.007953 | 0.010318 | 0.036457 | 0.007631 |
| | OF-Value | −0.3859 | −0.4243 | −0.3858 | −0.327 | −0.2503 | −0.4048 |
| | NS-Vectors | 34 | 21 | 35 | 37 | 38 | 27 |

models. It is not possible to plot all of them here, and we simply show the calibration, as well as the validation of GPR, applied to smoothing correction, and finally the comparative graphs of the validation of the three nonlinear regression models.

Here TM simply means the content of the paracetamol samples.

It is noticeable from the plot of the prediction of the three nonlinear calibrations, that the contents from KSVM (*T*-KSVM) and those from GPR (*T*-GPR) are very close. It is also visible that random forest regression is very bad when the content is small. This is surely an indication that random forest is over predicting the contents of the paracetamol when the content has low contents.

## 5. Conclusion

Nowadays, the NIR spectroscopy technique is undeniably one of the most easiest and practicable choices to perform data analysis such as control quality in the pharmaceutical world. In this work, we apply this non-destructive method to fifty-eight (58) 500 mg tablets of paracetamol of different origins in the local markets in Bamako. The regression models performed are Gaussian process regression, support vector machine, random forest, and partial least square regressions. The several parameters we compute, show that the best model is GPR applied to the smooth spectral correction of the original data. It is also demonstrated that random forest badly predicts contents of the paracetamol of much lower contents. From our analysis, one can surely conclude that GPR has the potential of quantifying paracetamol samples. Moreover, we witness that the three nonlinear regression techniques give reasonable statistical parameters compared to PLS. This will result in a huge impact on fighting against falsified samples of this antipyretic and painkillers product. We should stress that the limitation to fifty-eight different batch numbers is not a problem and will not alter our conclusion.

## Declaration of Competing Interest

The authors declare that they have no known competing financial interests or personal relationships that could have appeared to influence the work reported in this paper.

## Data availability

Data will be made available on request.

## Acknowledgments

The authors thank the members at LOSSA for useful discussions and also the members of the department of physics at USTTB-FST for their constant support and advice. Aminata SOW acknowledges financial support from SIDA (the Swedish International Development Cooperation Agency) through ISP (the International Science Program, Uppsala University). She is also grateful to the team at LNS where the UV part of this work has been conducted.